\documentclass{article}
\usepackage{times}
\usepackage{cite}
\usepackage{amsmath}
\usepackage[utf8]{inputenc}
\usepackage[english]{babel}

\usepackage{amsmath,amsfonts,amssymb,amsthm,epsfig,epstopdf,titling,url,array}

\theoremstyle{plain}

\theoremstyle{definition}
\newtheorem{defn}{Definition}[section]

\theoremstyle{remark}

\usepackage{graphicx}
\usepackage{latexsym}
\usepackage{subfigure}
\usepackage{pstricks}
\usepackage{setspace}
\begin{document}
\title{Application Specific Instrumentation (ASIN): A Bio-inspired Paradigm to Instrumentation using {\it recognition before detection}}
\author{
Amit Kumar Mishra\\
Electrical Engineering Department \\
University of Cape Town, South Africa\\
Email:akmishra@ieee.org
}
\date{}
\maketitle
\begin{abstract}
In this paper we present a new scheme for instrumentation, which has been inspired by the way small mammals sense their environment. 
We call this scheme Application Specific Instrumentation (ASIN). 
A conventional instrumentation system focuses on gathering as much information about the scene as possible. This, usually, is a generic system whose data can be used by another system to take a specific action. 
ASIN fuses these two steps into one. 
The major merit of the proposed scheme is that it uses low resolution sensors and much less computational overhead to give good performance for a highly specialised application. 

\end{abstract}
\maketitle
\section{Introduction}
\begin{figure}
\centering
	\includegraphics[width=90mm]{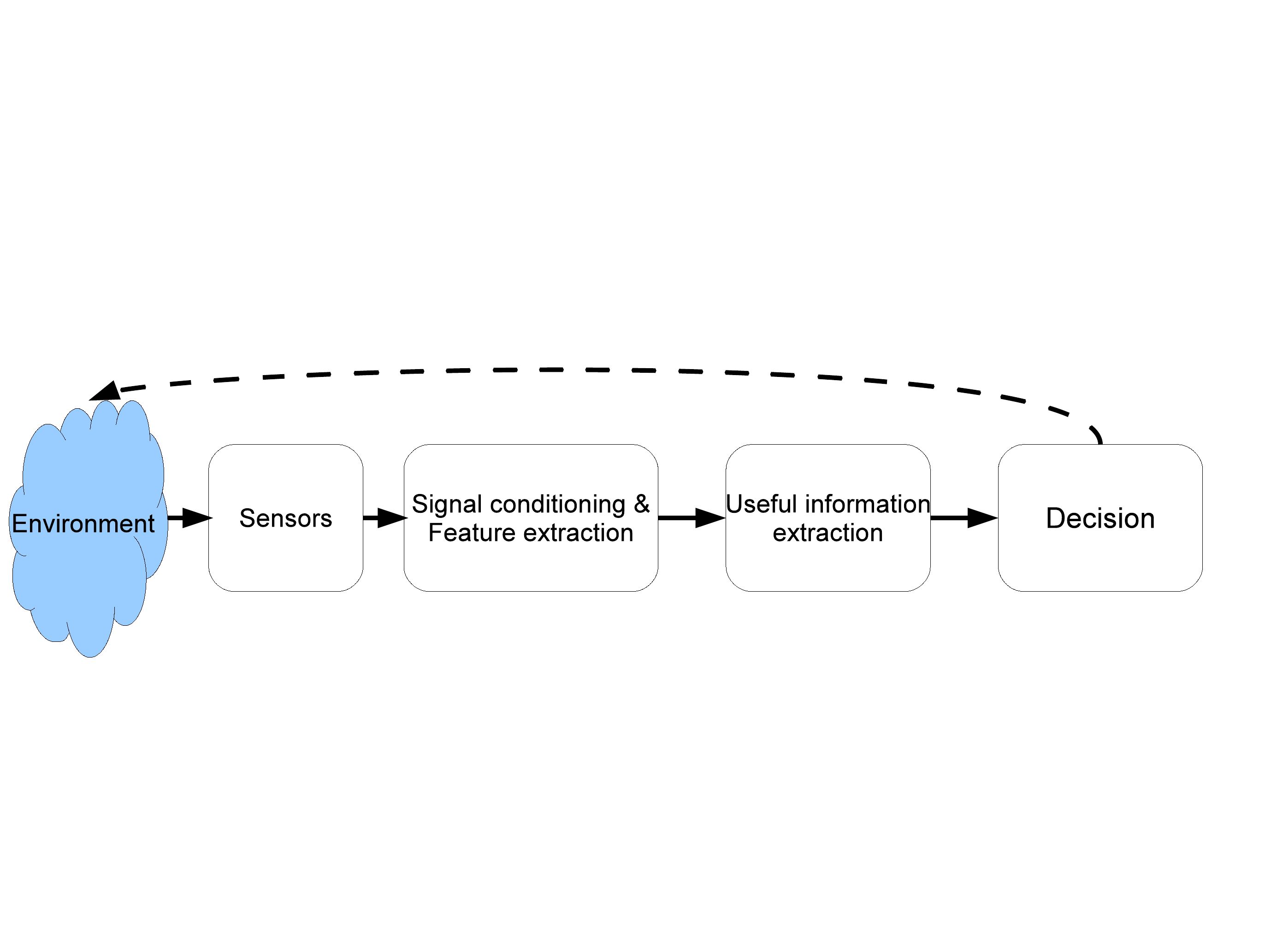}
	\caption{Chain of action blocks in conventional instrumentation systems} \label{chain}
\end{figure}

Conventional instrumentation systems have always aimed for higher sophistication to perform better. 
So the traditional instrumentation system will measure signal using high resolution sensors, then extract information from the signal, then interpret these and finally take some action based on the interpretation. 
The inherent philosophy in this has been the separation between sensor system and decision system. 
In other words sensors system are designed to be {\it generic}. 
The processing chain is further illustrated in Figure ~\ref{chain}. 
Mark that all the blocks after the $Sensor$ block can have human intervention and feedback. 
And the dashed link between the $Decision$ and $Sensor$ is advocated by most of the intelligent instrumentation works. 
The recently published work on intelligent and smart instruments \cite{pap_92,clarke_00} roughly follows the same chain of action with an added path for feedback from the decision making block to the sensor block.

\begin{figure}
\centering
	\includegraphics[width=90mm]{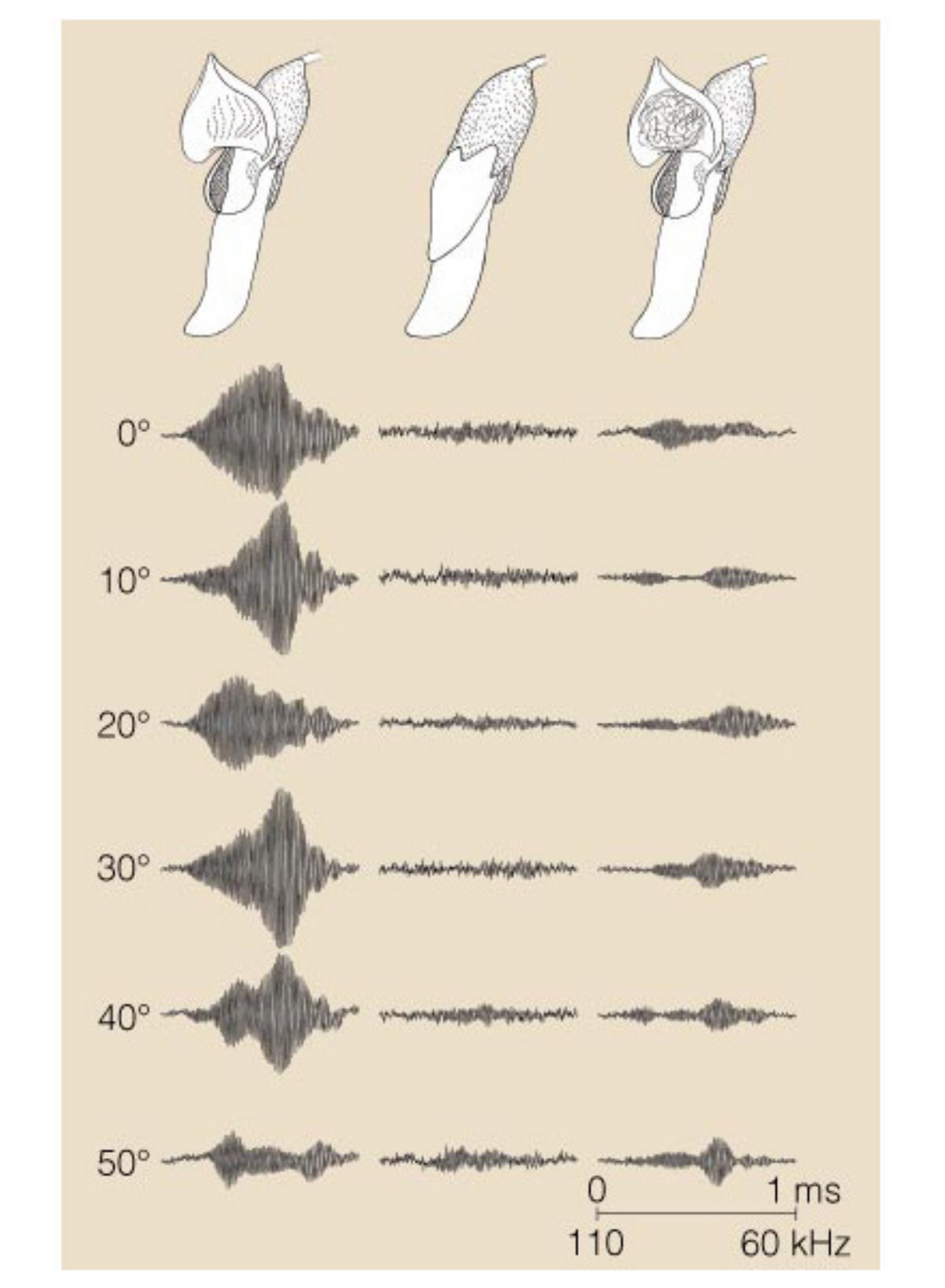}
	\caption{Pollen seeking bats' transmitted waves have evolved over generations to make them a rough information sensing system  \cite{von_99}.} \label{bats_evo}
\end{figure}

However, most living organisms follow a different approach in which they have evolved as a system for a very specific task. 
An example of this  is shown in Figure ~\ref{bats_evo} taken from reference \cite{von_99}. 
In this interesting figure, the scattered return of the sound waves generated by a type of pollen-feeding bat from different types of flowers (of the same species) are shown. 
As can be observed, the return is high (over a very wide angle of incidence) only if the flower is in full bloom and has real nectar in it! 
So, the sound wave generated by the bats have been fine-tuned over generations, so that the current generation can well be termed as information sensing processors. 
There are three major observations from this and other studies done on bats \footnote{The choice of bat is because of two reasons. First of all, because of its unique ability to use sound for visualizing its environment, there has been a lot of study on bat's sensors and brain. Secondly, being a small mammal these studies have been quite confirmatory in terms of how its brain works.}.
First of all, the sensing system has evolved for a very particular function. 
Secondly, there are multiple sensors that the bat uses during foraging. And none of these sensors are of high {\it resolution} in traditional meaning of the word. 
Lastly, bats brain is very simple and mostly works as a correlation processor \cite{zook_85}. It can also be noted here that extensive research on the working of visual cortex have also given similar conclusions. 
In other words even in evolved mammals like human brain does not process the sensory output as it is. Rather it always works in a goal-driven manner, using layers of neurone in prefrontal cortex  to perform certain predefined task. 


From these observations, we have worked on a new design of instrumentation, which we call application specific instrumentation (ASIN). 
We shall present some of our previous works where we have demonstrated the working of ASIN in various applications. 
The proposed scheme works on a {\it recognition before detection} philosophy. 
Hence, we shall also arrive at new definition for {\it resolution} in the context of ASIN. 

In rest of the paper I shall present the configuration of a generic ASIN system in Section II. 
In Section III a new framework for resolution shall be developed. Section IV will present some applications using ASIN. Lastly I shall conclude in the last section. 

\section{ASIN Configuration}
\begin{figure}
\centering
	\includegraphics[width=90mm]{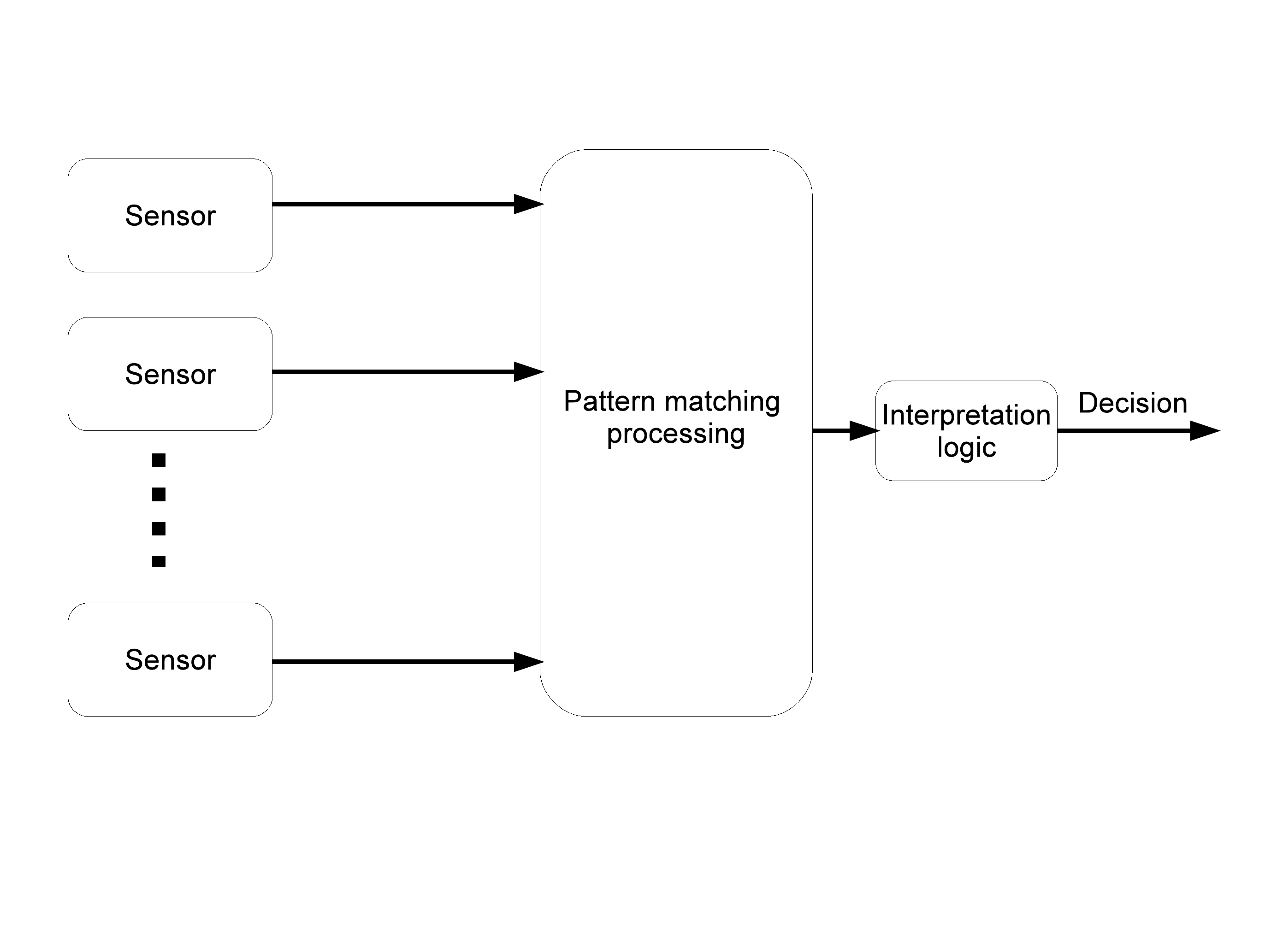}
	\caption{Block diagram of the basic ASIN instrument.} \label{asin_block}
\end{figure}

The  configuration of the proposed instrumentation is shown in Figure ~\ref{asin_block}.  
In the following, we describe the major components of ASIN system and how they work.
\begin{itemize}
\item {\it Sensors:} The sensors are the interface of ASIN to the environment. 
There are two major ways in which this is different from the sensor block of Figure ~\ref{chain}. 
First of all, each individual sensor in ASIN is of low or crude resolution. 
Secondly, each sensor interacts with the environment through a preselected application specific measurement matrix. 
Usually, if $e$ is the environment parameter to be sensed, a sensor measures $s$ through a measurement process $P$. 
I.e. $$ s = P e$$ 
A high resolution sensor is one which measures $e$ as truthfully as possible, and hence for an ideal sensor $P$ should be as close to an identity matrix as possible. 
The basic sensor is fine-tuned in ASIN with application specific measurement matrix, $P_{ASIN}$. 
Hence, the final measurement is $s'$, where $$s' = P_{ASIN} P e .$$
The design of a suitable application specific measurement matrix is one of the major tasks for ASIn design. 
This needs focusing the end-goal to as few decisions as possible and to try to make these decisions binary (to make the whole system simple). 
\item {\it Correlator Processor:} Given that the sensors have been designed with some intelligence, the next step is a simple correlation type processor, instead of a high performance processor (which usually handles signal processing and information extraction). 
\item {\it Interpreter:} The final block is Interpreter which interprets the output from the correlation processor and gives a final decision level. 
And as we discussed before, the final decision need to be as few as possible and preferably with a binary value. 
\end{itemize}

\section{Some Applications}
In this section, we describe two simulation based validation of a limited version of the ASIN system we discussed in the last section. 
The limitation of the version of ASIN used lies in the fact that the sensors used in this setup are generic instead of being customized for the final goal.
\subsection{ASIN based breast tumor detection}
In the first demonstration we use UWB based Radar to measure breast tumor.  
This is an active area of research \cite{davis_08,helbig_12}. 
But most of the work in this domain follow the conventional instrumentation path as elucidated in Figure ~\ref{chain}. 
The main focus is on the development of a generic instrument of high resolution. 
In the ASIN based approach, we fix the end goal first. We fixed the final goal to be the detection of the presence of tumor in breast. 
For this instead of generating a 3D image of the breast tissues, which requires the use of a UWB Radar array, we used just one UWB sensor node. 
The correlation processor block of Figure ~\ref{asin_block} was simulated using a single layer neural network, which was first trained with few training measurements (all using single UWB sensor). 
The final system was able to detect tumor with an accuracy of 98\% even in the presence of tissue bundles whose relative permitivity was much higher than that of normal tissue. 
Further details of the work can be found in \cite{mishra_10}.

\subsection{ASIN based sludge volume estimation}
\begin{figure}
\centering
	\includegraphics[width=50mm]{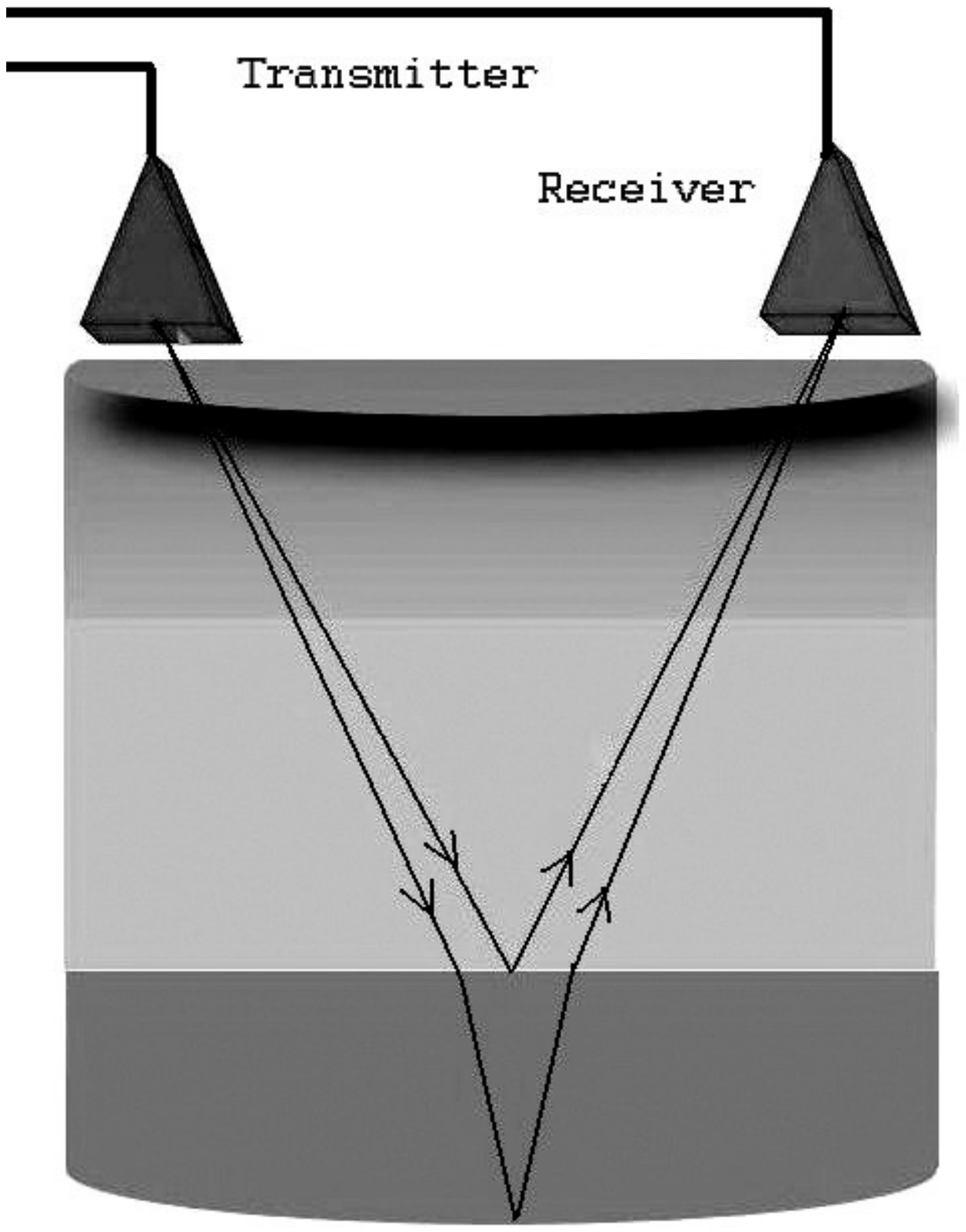}
	\caption{Setup for measuring sludge volume in oil tanks using a single UWB Radar sensor.} \label{uwb_tank}
\end{figure}

In a second attempt to implement an ASIN type system, we tried to build an instrument using UWB Radar to measure sludge volume in oil tanks. 
The experimental setup is shown in Figure ~\ref{uwb_tank}. 
Again, the conventional approach is to use an array of UWB transceiver sensor nodes to first form a 3D image of the tank bottom, and finally to use image processing steps to estimate the volume of sludge. 
In the ASIN based approach, we first trained the correlator processor, again implemented by a single layer neural network. 
In the testing phase, we changed the volume of sludge in incremental steps and for each volume we also changed the sludge profile. 
The performance of the system is best represented by a regression coefficient, between the actual volume and the estimated volume. 
We got a regression value of 0.91 using simulation based experiments, and of 0.88 using practical measurements. 

\section{Redefining Resolution}
One of the major figures of merit of any instrumentation system is its resolution. 
 As per the Measurement Systems Analysis Manual `` the resolution of an instrument is $\delta$ if there is an equal probability that the indicated value of any artifact, which differs from a reference standard by less than $\delta$, will be the same as the indicated value of the reference''. 
 There are few ways in which we will refine this definition to create a new definition of resolution for ASIN. 
Depending on that we shall define two different types of resolutions. 

\subsection{Neyman-Pearson Principle based Definition}
For a generic instrument, as it says in the definition of the resolution, they measure the ``value of any artifact''. In the context of ASIN this becomes the ``value of any artifact of interest''. Secondly the way we measure this artifact using ASIN is by using pattern recognition algorithms. 
Hence the output from ASIN is mostly a propbability, $P_p$ that a certain ``artifact of interest'' is present. 
Let us represent the artifact of interest as $\theta$.
For a given task there will always be cases of misclassification which will create events of false alarm giving a probability of false present, $P_{fp}$. 
Then using Neyman-Pearson principle the goal of desinging an ASIN is to maximize $P_p$ for a given $P_{fp}$. 

\begin{defn}
The Neyman-Pearson principle based resolution ($\delta_{NP}$) can be defined as  the maximum change in the artifact of interest, $\Delta \theta$, which for a given ASIN makes sure that the change in the probability of false present $\Delta P_{fp} \leq$ an arbitrary small number $\epsilon$.
\end{defn}

\subsection{Cramer-Rao Bound based Definition}
From a data-centric point of view for an ``artifact of interest'', $\theta$, we use ASIN system to capture some signal $\zeta$. 
One of the major empowering concept of ASIN is the hypothesis that we can estimate  $\theta$ from $\zeta$ even when the exact phenomenological link from  $\theta$ to $\zeta$ is not well-modeled. 
This way of putting the ASIN operation helps in looking at the problem from a parameter-estimation point of view. 
And in parameter estimation one of the important figures of merit is Cramer-Rao lower bound (CRLB) which is the lowest variance in the estimate. 
CRLB is the inverse of Fisher information\footnote{As we are discussing about resolution with respect to any given artifact of interest we will only deal with scalar $\theta$.}  $$I(\theta) = E\left[  \left( \frac{\partial l(\zeta;\theta)}{\partial \theta}  \right)^2 \right]$$, where $l(\zeta;\theta)$ is the natural logarithm of the likelihood function and $E$ represents the expectation operation. 
For ill-defined mapping like what we deal with in ASIN Fisher information can be found numerically as well \cite{spall_12}. 

\begin{defn}
The Cramer-Rao principle based resolution ($\delta_{CR}$) can be defined as  the maximum change in the artifact of interest, $\Delta \theta$, which for a given ASIN makes sure that the change in the Fisher information $\Delta I(\theta) \leq$ an arbitrary small number $\epsilon$.
\end{defn}

\section{Conclusion and Discussion}
We propose a new way to instrumentation which we name as application specific instrumentation (ASIN). 
This is a biologically inspired scheme, where the focus is on designing instruments for very focused problems, so focused that the decision can be made binary. 
The proposed scheme is expected to be less costly, require much less computational overhead and to perform better for specialized applications. 
We have demonstrated the scheme in a limited sense for two usages involving ultrawideband Radars. 
In both these we have shown that the end goal has been achieved using a single pair of transmitter-receiver instead of an array. 
One of the important steps of the whole scheme is the design of the application specific measurement matrix. 

We also present some modified  definitions of resolution which can be used to quantify the performance of an ASIN system.

\bibliographystyle{IEEEtran}
\bibliography{asin_bib}

%
%

\end{document}